\documentstyle[12pt]{article}

\def\hybrid{\topmargin -20pt  \oddsidemargin 0pt
      \headheight 0pt   \headsep 0pt
      \textwidth 6.25in 
      \textheight 9.5in 
      \marginparwidth .875in
      \parskip 5pt plus 1pt   \jot = 1.5ex}

\hybrid

\begin{document}
\def\x{\times}
\def\beq{\begin{equation}}
\def\eeq{\end{equation}}
\def\beqa{\begin{eqnarray}}
\def\eeqa{\end{eqnarray}}

\sloppy
\newcommand{\be}{\begin{equation}}
\newcommand{\eq}{\end{equation}}
\newcommand{\ov}{\overline}
\newcommand{\un}{\underline}
\newcommand{\p}{\partial}
\newcommand{\la}{\langle}
\newcommand{\ra}{\rangle}
\newcommand{\bl}{\boldmath}
\newcommand{\ds}{\displaystyle}
\newcommand{\nl}{\newline}
\newcommand{\th}{\theta}



\renewcommand{\thesection}{\arabic{section}}
\renewcommand{\theequation}{\thesection.\arabic{equation}}

\parindent1em


\begin{titlepage}
\begin{center}
\hfill HUB-EP-97/32\\
\hfill {\tt hep-th/9705197}\\

\vskip .7in

{\bf  Chiral Multiplets in $N=1$ Dual String Pairs}

\vskip .3in

Gottfried Curio\footnote{email: curio@qft2.physik.hu-berlin.de}
\\
\vskip 1.2cm

{\em Humboldt-Universit\"at zu Berlin,
Institut f\"ur Physik, 
D-10115 Berlin, Germany}

\vskip .1in

\end{center}

\vskip .2in

\begin{center} \end{center}
\begin{quotation}\noindent

We compare the spectrum of chiral multiplets in the $N=1$ vacua
of the heterotic string on a Calabi-Yau together with an
$E_8\times E_8$ vector bundle and $F$-theory on a smooth Calabi-Yau
fourfold. Under suitable restrictions we show agreement
using an index-computation.

\end{quotation}
\end{titlepage}
\vfill
\eject

\newpage

Two of the most interesting classes of models coming from string theory
compactifications with $N=1$ supersymmetry in four dimensions are obtained
from the heterotic string on a threedimensional Calabi-Yau $Z$ with a
vector bundle $V$ respectively from $F$-theory on an elliptically
fibered Calabi-Yau fourfold $X^4 \rightarrow B^3$. These cases are 
of even greater 
interest in view of the duality philosophy that one should get dual models
by adiabatically extending the eightdimensional duality between the heterotic
string on $T^2$ and $F$-theory on $K3$ [\ref{V}]. 
So we will furthermore assume
that the heterotic Calabi-Yau is elliptically fibered $Z \rightarrow B^2$
and that the $F$-theory fourfold Calabi-Yau is $K3$ fibered 
$X^4 \rightarrow B^2$. This involves also the existence of a $P^1$ fibration
$B^3 \rightarrow B^2$. \\

In this note we will show a duality matching 
among the concerned moduli assuming that we are
in the case without an unbroken gauge group. So we will assume that $V$ is an
$E_8 \times E_8$ vector bundle and that $X^4$ is a smooth Weierstrass model;
for worked out examples cf. [\ref{ACL}].
The comparison works technically similar to the matching [\ref{FMW}] of the 
number $\chi /24$
of spacetime-filling threebranes [\ref{SVW}] occuring in the course of
tadpole-cancellation to the number $h$ of fivebranes (wrapping the 
fibre $F$ of the elliptic fibration $Z \rightarrow B^2$) 
occuring in the context of the heterotic string anomaly cancellation

\beqa
\lambda (V_1)+\lambda (V_2)+h [F]=c_2(Z)\nonumber
\eeqa
in connection with a $E_8 \times E_8$ vector bundle $V=V_1 \times V_2$. Note 
also that at the end of this paper we remark on circumstances that the actual
computation carried out here will find its proper place when embedded in
an enlarged context.

 So we consider $F$-theory on a smooth elliptically fibered Calabi-Yau 
fourfold $X^4$ with base $B^{3}$ which can be represented 
by a smooth Weierstrass model. We assume even more that the rank $v$ of the
gauge group in four dimensions is zero (,i.e. also no $U(1)$ factors).
Now one can expect that the spectrum would get contributions from
Kaehler and complex structure parameters related to $h^{1,1}-2$
(not counting the unphysical $F$-theory elliptic fibre as well as not
counting the class corresponding to the heterotic dilaton) and $h^{3,1}$
respectively as well as from $h^{2,1}$ giving in total
$h^{1,1}-2+h^{2,1}+h^{3,1}$ parameters.
One can see from a $M$-theory versus $F$-theory consideration that 
these contributions divide
themselves in 4$D$ between chiral and vector multiplets according to
whether or not they come from the threefold base $B^3$ of the $F$-theory
elliptic fibration (just as in the analogous 6$D$ $N=2$ case [\ref{MV2}]; we 
will assume here $v=0$ anyway).

So following this line of reasoning one would 
expect for the rank $v$ of the $N=1$ vector multiplets (unspecified
hodge numbers relate to $X^4$) (cf. [\ref{Mo}],[\ref{CL}],[\ref{ACL}])

\beqa
v=h^{1,1}-h^{1,1}(B^3)-1+h^{2,1}(B^3)\nonumber 
\eeqa
and for the number $c$ of $N=1$ neutral chiral 
(resp. anti-chiral) multiplets

\beqa
c&=&h^{1,1}(B^3)-1+h^{2,1}-h^{2,1}(B^3)+h^{3,1}\nonumber\\
 &=&h^{1,1}-2+h^{2,1}+h^{3,1}-v\nonumber
\eeqa
 
Note that we are speaking here of the generic gauge group and not of
the situation where one tunes the deformations of the Calabi-Yau to
unhiggs an unbroken gauge group and correspondingly counts only the
number of deformations {\em preserving} a particular singular locus
(cf. [\ref{BJPS}]).\\
 
Now one has [\ref{SVW}]

\beqa
\frac{\chi}{6}-8=h^{1,1}-h^{2,1}+h^{3,1}\nonumber
\eeqa
so that one finds

\beqa
c=\frac{\chi}{6}-10+2h^{2,1}-v\nonumber
\eeqa
Now one can compute the Euler number of X in terms of topological 
data of the base $B^3$ of the elliptic fibration according to [\ref{SVW}]

\begin{eqnarray}
\frac{\chi}{24}=12+15\int_{B^3}c_{1}^3(B^3)\nonumber
\end{eqnarray}

One can go even further: assuming, in the light of
the application to duality with the heterotic string, 
that $X^4$ is a $K3$ fibration over a twofold base $B^2$ (so that
$B^3$ is a $P^1$ fibration over $B^2$) one gets [\ref{FMW}]

\begin{eqnarray}
\frac{\chi}{24}=12+90\int_{B^2}c_{1}^2(B^2) +30\int_{B^2} t^2\nonumber
\end{eqnarray}
where $t$ encodes the $P^1$ fibration structure [\ref{FMW}]: 
assume the $P^1$ bundle
over $B^2$ given by the projectivization of a vector bundle 
$Y={\cal O}\oplus {\cal T}$, with ${\cal T}$ a line bundle over $B^2$; then
$t=c_1({\cal T})$. So one gets in the case with no unbroken gauge group

\begin{eqnarray}
c_F=\frac{\chi}{6}-10+2 h^{2,1}=
38+360\int_{B^2}c_{1}^2(B^2)+120\int_{B^2} t^2+2h^{2,1}\nonumber 
\end{eqnarray}

On the other hand we have to count the moduli on the heterotic side. Here
one has contributions $m_{geo}=h^{1,1}(Z)+h^{2,1}(Z)$ from geometrical moduli
plus the bundle moduli. Concerning the former
let us assume that the dual heterotic Calabi-Yau
threefold $Z$ is smooth so that [\ref{Klemm}]

\begin{eqnarray}
\chi (Z)=-60\int_{B^2}c_1^2(B^2)\nonumber
\end{eqnarray}
and assume furthermore that its smooth Weierstrass model is general
[\ref{Gr}], i.e. has only one section (a typical counterexample being
the $CY^{19,19}=B_9\times _{P^1}B_9$ with the del Pezzo $B_9$, the blow up
of $P^2$ in the nine points of intersection of two cubics, cf. [\ref{DGW}],
[\ref{CL}]). So
$h^{1,1}(Z)=h^{1,1}(B^2)+1$; this gives using Noethers formula that

\begin{eqnarray}
h^{1,1}(Z)=\int_{B^2}c_2(B^2)-1=11-\int_{B^2}c_1(B^2)^2\nonumber 
\end{eqnarray}
So one gets for the number of geometrical moduli

\begin{eqnarray}
m_{geo}&=&h^{1,1}(Z)+h^{2,1}(Z)=2 h^{1,1}(Z)-\frac{\chi}{2}\nonumber\\
       &=&22+28\int_{B^2}c_1^2(B^2)\nonumber
\end{eqnarray}

Concerning the contribution of the bundle moduli of the $E_8 \times E_8$
bundle $V$ on $Z$ let us recall the
setup of the index-computation in [\ref{FMW}]. As the usual quantity
suitable for index-computation $\sum_{i=o}^3 (-1)^i h^i(Z,V)$ vanishes
by Serre duality one has to introduce a further twist and to compute a
character-valued index. Now because of the elliptic fibration structure
one has on $Z$ the involution $\tau$ coming from the "sign-flip" in the fibers
which we furthermore assume has been lifted to an action on the bundle.
The character-valued index

\begin{eqnarray}
I=-\frac{1}{2}\sum_{i=o}^3 (-1)^i Tr_{H^i(Z,V)} \tau \nonumber
\end{eqnarray}
simplifies by the vanishing of the ordinary index to

\begin{eqnarray}
I=-\sum_{i=o}^3 (-1)^i h^i(Z,V)_e \nonumber
\end{eqnarray}
where the subscript "e" (resp. "o") indicates the even (resp. odd) part. As
we have the gauge group completely broken one finds 

\begin{eqnarray}
I=n_e-n_o \nonumber
\end{eqnarray}
denoting by $n_{e/o}$  the number $h^1(Z,V)_{e/o}$ of massless even/odd
chiral superfields.
Now one gets with Wittens index formula [\ref{W-un}],[\ref{ACL}]

\begin{eqnarray}
I=16+332\int_{B^2}c_1^2(B^2) +120\int_{B^2} t^2 \nonumber
\end{eqnarray}
that one has for the number of the bundle moduli $m_{bun}=n_e+n_o=I+2n_o$

\begin{eqnarray}
m_{bun}=16+332\int_{B^2}c_1^2(B^2) +120\int_{B^2} t^2 +2n_o \nonumber
\end{eqnarray}
So adding up one gets in total

\begin{eqnarray}
c_{het}&=&h^{1,1}(Z)+h^{2,1}(Z)+I+2n_o\nonumber\\
       &=&38+360\int_{B^2}c_1^2(B^2)+120\int_{B^2} t^2+2n_o\nonumber
\end{eqnarray}

Now on the $F$-theory side the modes odd under the involution $\tau ^{\prime}$
corresponding to the heterotic involution $\tau$ correspond 
to the $h^{2,1}(X^4)$ classes [\ref{FMW}]. Let us assume that 
no 4-flux was turned on (which is not a free decision [\ref{W4fl}] 
in general); otherwise there are further twistings possible (cf. [\ref{FMW}],
sect. 4.4, and also [\ref{AM}]) which account for a possible multi-component
structure of the bundle moduli space (cf. also [\ref{BJPS}] for the case of
$SU(n)$ bundles). This clearly deserves further study to embed the simple
picture employed here in the more general context. \\ 
So one gets complete matching with $n_o=h^{2,1}(X^4)$.\\
 
I would like to thank B. Andreas and E. Witten for discussion.

\section*{References}
\begin{enumerate}

\item
\label{V}
C. Vafa, {\it Evidence for F-theory}, Nucl. Phys. {\bf B 469} (1996) 493,
hep-th/9602022.

\item
\label{FMW}
R. Friedman, J. Morgan and E. Witten, {\it Vector Bundles and 
F-Theory}, hep-th/9701162.

\item
\label{SVW}
S. Sethi, C. Vafa and E. Witten, {\it Constraints on Low-Dimensional String
Compactifications}, Nucl. Phys. {\bf B 480} (1996) 213, hep-th/9606122.

\item
\label{MV2}
D. R. Morrison and C. Vafa, {\em Compactification of F-theory on 
Calabi-Yau Threefolds II}, Nucl. Phys. {\bf B 476} (1996) 437, hep-th/9603161.

\item
\label{Gr}
A. Grassi, {\it Divisors on elliptic Calabi-Yau 4-folds and the superpotential
in F-theory}, alg-geom/9704008.

\item
\label{Mo}
K. Mohri, {\it F- Theory Vacua in Four Dimensions And Toric Threefolds},
hep-th/9701147.

\item
\label{Klemm}
A. Klemm, B. Lian, S.-S. Roan and S.-T. Yau, {\it Calabi-Yau fourfolds
for $M$- and $F$-Theory Compactifications}, hep-th/9701023.

\item
\label{W-un}
E. Witten, unpublished notes.

\item
\label{W4fl}
E. Witten, {\it On Flux Quantization in M Theory and the Effective Action},
hep-th/9609122.

\item
\label{DGW}
R. Donagi, A. Grassi and E. Witten, {\it A Nonperturbative Superpotential with
$E_8$ Superpotential}, Mod. Phys. Lett. {\bf A 11} (1996) 2199, hep-th/9607091.

\item
\label{CL}
G. Curio and D. L\"ust, {\it A Class of $N=1$ Dual String Pairs and its
Modular Superpotential}, to appear in Int. Journ. of Mod. Phys. {\bf A},
hep-th/9703007.

\item
\label{ACL}
B. Andreas, G. Curio and D. L\"ust, {\it $N=1$ Dual String Pairs and their
Massless Spectra}, hep-th/9705174.

\item
\label{AM}
P.S. Aspinwall and D.R. Morrison, {\em Pointlike Instantons on K3 Orbifolds},
hep-th/9705104.

\item
\label{BJPS}
M. Bershadsky, A. Johansen, T. Pantev and V. Sadov, {\it On Four-Dimensional 
Compactifications of F-Theory}, hep-th/9701165.

\end{enumerate}

\end{document}